\PassOptionsToPackage{svgnames}{xcolor}

\documentclass[sigconf,nonacm]{acmart}

\usepackage[T1]{fontenc}        
\usepackage{hyperref}           
\usepackage{cleveref}           
\usepackage{url}                
\usepackage{booktabs}           
\usepackage{amsfonts}           
\usepackage{nicefrac}           
\usepackage{microtype}          
\usepackage{subcaption}         
\usepackage[normalem]{ulem}     
\usepackage{lipsum}
\usepackage{multirow}

\setcopyright{none}
\copyrightyear{2023}
\acmYear{2023}

\title{Energy Efficiency of Quantum Statevector Simulation at Scale}

\author{Jakub Adamski}
\email{jakub.adamski@ed.ac.uk}
\orcid{0009-0000-5221-2070}
\affiliation{%
    \institution{EPCC, The University of Edinburgh}
    \city{Edinburgh}
    \country{UK}
}

\author{James Peter Richings}
\email{j.richings@epcc.ed.ac.uk}
\orcid{0000-0003-0895-4063}
\affiliation{%
    \institution{EPCC, The University of Edinburgh}
    \city{Edinburgh}
    \country{UK}
}

\author{Oliver Thomson Brown}
\email{o.brown@epcc.ed.ac.uk}
\orcid{0000-0002-5193-8635}
\affiliation{%
    \institution{EPCC, The University of Edinburgh}
    \city{Edinburgh}
    \country{UK}
}

\renewcommand{\shortauthors}{Adamski, Richings, and Brown}

\ccsdesc[500]{Hardware~Impact on the environment}
\ccsdesc[100]{Applied computing~Physics}
\ccsdesc[300]{Theory of computation~Quantum computation theory}
\ccsdesc[300]{Computing methodologies~Parallel computing methodologies}

\begin{document}

\begin{abstract}
Classical simulations are essential for the development of quantum computing, and their exponential scaling can easily fill any modern supercomputer. In this paper we consider the performance and energy consumption of large Quantum Fourier Transform~(QFT) simulations run on ARCHER2, the UK's National Supercomputing Service, with QuEST toolkit. We take into account CPU clock frequency and node memory size, and use cache-blocking to rearrange the circuit, which minimises communications. We find that using 2.00~GHz instead of 2.25~GHz can save as much as 25\%~of energy at 5\%~increase in runtime. Higher node memory also has the potential to be more efficient, and cost the user fewer CUs, but at higher runtime penalty. Finally, we present a cache-blocking QFT circuit, which halves the required communication. All our optimisations combined result in 40\%~faster simulations and 35\%~energy savings in 44~qubit simulations on 4,096~ARCHER2 nodes.  \\
\\
\copyright Adamski, Richings, Brown 2023. This is the author's version of the work. It is posted here for your personal use. Not for redistribution. The definitive version was published in Workshops of The International Conference on High Performance Computing, Network, Storage, and Analysis (SC-W 2023), \url{https://doi.org/10.1145/3624062.3624270}.
\end{abstract}

\maketitle

\section{Introduction}
\label{sec:introduction}

Quantum computing has become a major area of research in recent years. In 2019, Google published a paper, in which they claimed to have successfully performed the classically intractable task of random circuit sampling~\cite{Arute19}, leading to an explosion of public interest and investment in quantum computers. The technology is still in its infancy and noise is still a major challenge for quantum hardware, so classical simulation plays an important role in algorithm development and testing. It is straightforward to fill any known or planned supercomputer with exact simulations of fewer than 50~qubits. Given the rapidly growing interest in these unprecedentedly large simulations, we believe it is important to consider their sustainability. ARCHER2 has been used to simulate 44-qubit systems using 4,096~nodes with 1~PB of memory\footnote{A feat in part inspired by a recent AWS HPC blog post where the authors also ran a 44~qubit simulation using QuEST: \url{https://aws.amazon.com/blogs/hpc/simulating-44-qubit-quantum-circuits-using-aws-parallelcluster/}.} -- a number far below the size of the recently announced 433-qubit `Osprey' quantum processor from IBM\footnote{\url{https://newsroom.ibm.com/2022-11-09-IBM-Unveils-400-Qubit-Plus-Quantum-Processor-and-Next-Generation-IBM-Quantum-System-Two}.}. This application domain is growing in popularity and often requires use of most of the available resources on a given target hardware, entailing high energy consumption. Focusing on its efficiency therefore plays a crucial role in HPC sustainability. 

Most frequently, classical simulations are implemented by the Schr\"{o}dinger's algorithm, which is a na\"{i}ve but exact approach to simulating quantum circuits. The principle is to keep the whole statevector in memory, and evolve it by applying gates (a matrix-vector product) from the simulated circuit. This requires keeping track of all $2^n$ complex amplitudes, for $n$ qubits, and modifying them with each iteration. The maximum number of qubits in the register is bound by the total memory available on the target hardware. At the same time, the runtime of a circuit simulation scales linearly with the number of gates in the circuit, for a fixed size quantum register.

The main advantage of the statevector approach is that once a circuit is simulated, all amplitudes are available, which enables any required measurements to be made without the need to rerun the simulation. In contrast, many alternative approaches, like the Feynman path algorithm~\cite{Aaronson2016}, or tensor network contraction~\cite{Pan2022}, are restricted to a limited number of amplitudes in the output state in order to limit the memory usage. The trade-off is that those methods typically require repeated runs to perform all the desired measurements. This is also true of real quantum hardware, where the output of an individual run is always a classical bit string.

\section{Methodology}
\label{sec:methodology}

\subsection{QuEST Software}
\label{sec:methodology/quest-software}

QuEST \cite{Jones2019} is a \textit{statevector simulator} implemented in C and parallelised with MPI+OpenMP, which acts as described in \cref{sec:introduction}. In order to evolve the statevector, each amplitude needs to be updated for every applied gate. 

We can distinguish three kinds of quantum operators:
\begin{itemize}
    \item \textit{Fully local gates} -- each amplitude can be updated without accessing other amplitudes. Mathematically, these are diagonal matrices. 
    \item \textit{Local memory gates} -- each update is a linear combination of amplitudes on the same process. As matrices, those gates are block-diagonal, with blocks size less-or-equal to the number of amplitudes per process. 
    \item \textit{Distributed gates} -- new amplitudes depend on amplitudes from other processes. 
\end{itemize}

As long as the update dependencies are located on the same node, no message passing is required, and the gates can be applied relatively quickly. When the operators become distributed, MPI takes up a large portion of the runtime. Hence, distributed gates are much more expensive to execute. 

QuEST requires the statevector to be split evenly across $2^n$ processes. This ensures pairwise communication for any given gate. It also means that the entire local statevector needs to be exchanged -- 64~GB per process on ARCHER2 assuming one MPI process per node. This was the case in all experiments presented here. Due to limitations of some implementations of MPI, individual messages cannot be larger than 2~GB, so the communication cannot be done in a single message. Instead, 32~messages are exchanged per distributed gate, and QuEST implements the communications as a sequence of blocking \texttt{MPI\_Sendrecv}.

\subsection{Potential Optimisations}
\label{sec:methodology/optimisations}

We have identified three different strategies that help perform statevector simulations more efficiently. 

\begin{enumerate}
    \item \textbf{Changing the clock frequency of the CPU. } ARCHER2 allows users to set the CPU frequency of computing nodes to reduce energy consumption. The default currently is 2.00~GHz (medium), but users have the ability to change this to either 1.50~GHz (low), or 2.25~GHz (high) with a SLURM environment parameter\footnote{\url{https://docs.archer2.ac.uk/user-guide/energy/\#controlling-cpu-frequency}.}.
    
    \item \textbf{Using nodes with larger memory size. } There are two node types available on ARCHER2 -- standard (256~GB), and high memory nodes (512~GB). We can use fewer high-mem nodes for a given size state vector simulation. 
    
    \item \textbf{Transpiling the circuit to reduce communication via cache-blocking. } \textit{Cache-blocking} is a technique that can improve memory locality of the circuit. When a gate acts on lower index qubits, the dependencies between statevector elements are located closer together. It is possible to lower a target qubit index by swapping it with another qubit. Each distributed SWAP gate requires communication, but it can be compensated if the target is frequently acted on. Although not implemented in QuEST, cache-blocking is already a part of some frameworks, such as Qiskit, where it serves as a means of distribution to multiple processes~\cite{Doi2020}, and cuQuantum for multi-GPU support~\cite{Faj2023}. However, it is the first time this optimisation is used at such a scale and for energy reasons. 
    
\end{enumerate}

\subsection{Circuits}
\label{sec:methodology/circuits}

\begin{figure}[ht]
    \centering
    \begin{subfigure}{.45\textwidth}
        \centering
        \includegraphics[width=.75\textwidth]{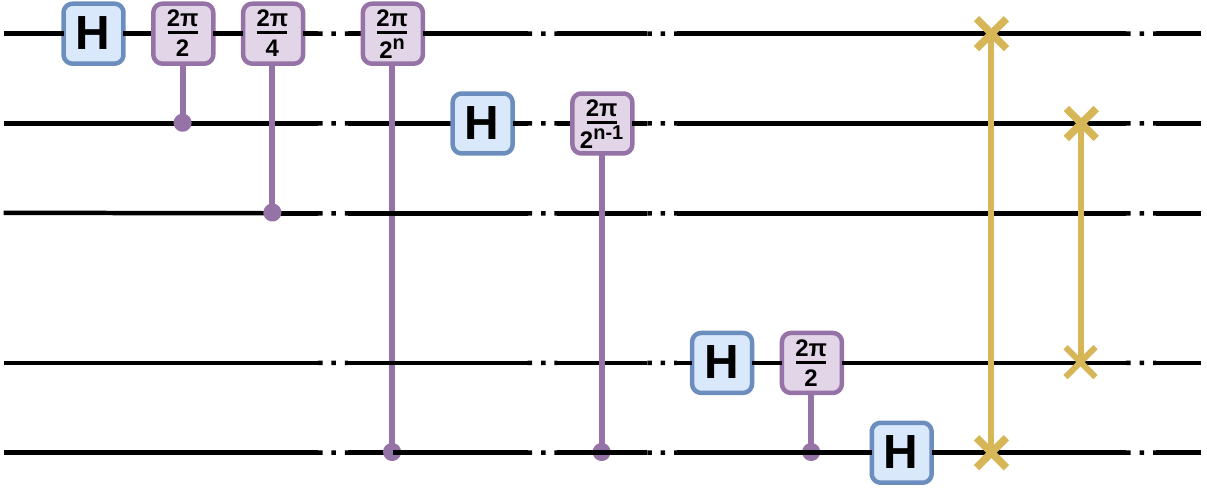}
        \caption{Standard.}
        \label{fig:qft-standard}
    \end{subfigure}
    \begin{subfigure}{.45\textwidth}
        \centering
        \includegraphics[width=.75\textwidth]{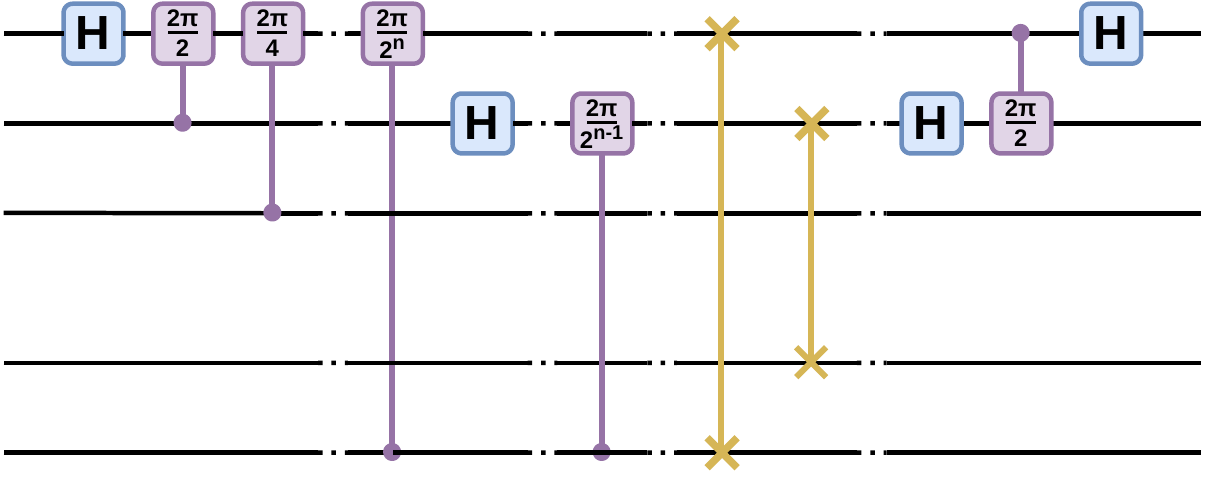}
        \caption{Cache-blocked.}
        \label{fig:qft-cache-blocking}
    \end{subfigure}
    \caption{Standard and cache-blocked QFT circuit.}
    \label{fig:qft-comparison}
\end{figure}

The main circuit used in this paper is the \textit{Quantum Fourier Transform (QFT)}, shown on \cref{fig:qft-standard}. It is a common subroutine of larger quantum algorithms, like Quantum Phase Estimation.

The QFT is straightforward to cache-block without adding new gates, as it already features SWAP gates at the end. By shifting them to the left, we can guarantee that all Hadamard gates are applied to local qubits. Gates to the right of the swaps need to be vertically flipped. As a result, the only communication remains in the distributed SWAP gates. An example is shown in \cref{fig:qft-cache-blocking}, where the last two Hadamard gates were made local. 

To test the effects of communication, two other benchmarking circuits were designed -- the Hadamard gate benchmark and the SWAP gate benchmark. Their structure is simple, consisting of $k$ gates applied sequentially to the same target qubits. Despite lack of practical use, they give insight into properties of statevector simulations. Importantly, a Hadamard benchmark on the last qubit of the system should be the worst-case simulation scenario, since all gates in the circuit are guaranteed to be distributed (provided the simulation spans multiple nodes).

\subsection{Energy Measurements and Profiling}
\label{sec:methodology/measurements}

The energy consumption of statevector simulations can be retrieved by querying SLURM on ARCHER2, which uses power counters on the nodes to determine energy usage. However, this is missing the energy used by the network, which we estimated with the following formula: \( E_{net} = n_s \cdot \bar{P}_s \cdot \Delta t \). The value $n_s$ represents the number of switches in the utilised network -- 1~switch per 8~nodes on ARCHER2. The typical average power of a switch under load on ARCHER2 $\bar{P}_s$ is equal to 235~W. Finally, $\Delta t$ is the total runtime of the circuit. The estimated switch energy and the energy extracted from SLURM are added together to calculate the total energy consumed by the jobs being studied. We have omitted other sources of energy consumption during the job, as we do not have values for other rack level overheads nor do we include energy consumed in cooling the system.

\section{Results}

We ran a QFT circuit at register sizes from 33 to 44 qubits, using the minimum possible number of nodes to fit the statevector, as well as the Hadamard and SWAP circuits to calculate the cost of distributed gates\footnote{Resources used are available at \url{https://github.com/jjacobx/arb23}.}.

\subsection{Clock Frequency and Node Memory}

\begin{figure}[ht]
    \centering
    \includegraphics[width=.8\linewidth]{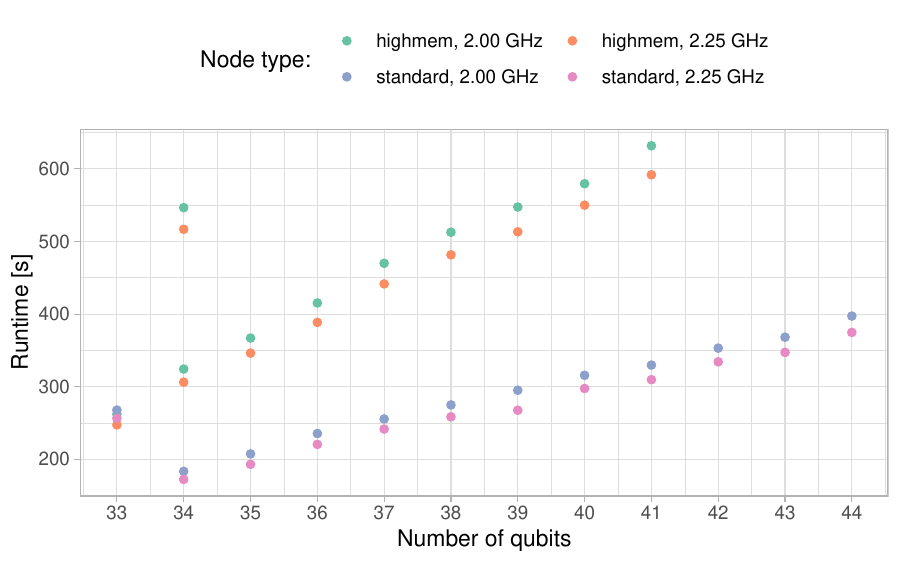}
    \caption{Runtimes of the QFT circuit simulations with different register size. Standard nodes have 256~GB, highmem nodes have 512~GB.}
    \label{fig:runtime-mem-freq-qft}
\end{figure}

To benchmark the most efficient combination of the CPU frequency and available memory, we ran the QFT circuit. Results are shown in \cref{fig:runtime-mem-freq-qft}. Additional buffers are required in the MPI implementation, doubling the overall memory requirement. As a result, 33~qubits will fit on a standard node, but 4~nodes are required for a 34~qubit simulation. The 33~qubit standard node results, and slower of the two 34~qubit high memory node results are single node runs.

QFT runtimes scale linearly, due to the number of distributed gates rising linearly. The actual number of gates grows quadratically. High memory nodes are slower, but less than twice as slow as the standard nodes. Higher memory means twice as much data in the local memory needs to be recomputed at each gate application. A maximum of 41~qubits could be simulated on 256~high memory nodes, and 44~qubits on 4,096~standard nodes. 

\begin{figure}[ht]
    \centering
    \includegraphics[width=.8\linewidth]{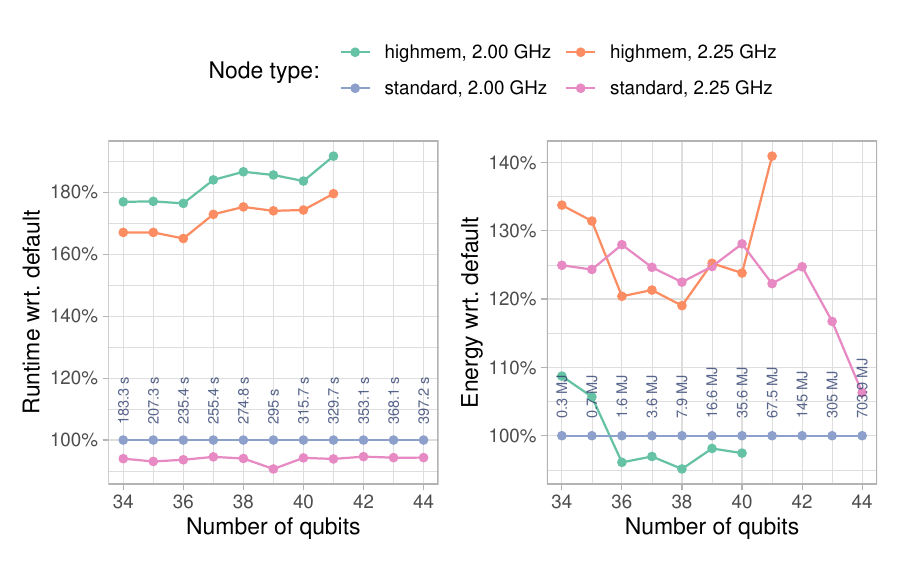}
    \caption{Comparison of the default medium frequency, standard nodes to other setups.}
    \label{fig:runtime-energy-percs-qft}
\end{figure}

Since the standard node medium frequency setup is currently the default on ARCHER2, we plotted its fractional comparison to other setups on \cref{fig:runtime-energy-percs-qft}, in terms of runtime and energy consumption. By interpreting both plots together, it is possible to assess which configuration is most optimal. The standard high frequency setup is consistently 5\%~to 10\%~faster than the default, but it uses around 25\%~more energy, up to 43~qubits, where the consumption significantly decreases. Therefore, we conclude that the defaults are appropriate for most simulations.

The use of high memory nodes drastically increases the runtime, however the CU cost of high memory simulations is lower than for standard memory. High frequency on high memory nodes required from 20\%~to over 40\%~more energy. We found that the lowest frequency available on ARCHER2 (1.5~GHz) was not of benefit in either case due to a large increase in runtime, and so these runs are omitted from our figures.

\subsection{Cache Blocking}

The cache blocking approach was described in \cref{sec:methodology/optimisations}. In addition to the QFT, we ran the two other benchmarks from~\cref{sec:methodology/circuits}, with 50~gates in each circuit. During the investigation, we noted that QuEST uses a series of blocking communications. We rewrote the message passing function to use non-blocking communication, which allows multiple messages to be sent and received in parallel when using an interconnect with high bandwidth. Each of our experiments here includes results from both the default QuEST implementation and our modified version.

\begin{table}[ht]
    \centering
    \begin{tabular}{r | c c | c c}
        \multirow{2}{*}{Qubit Index} & \multicolumn{2}{c|}{Blocking}  & \multicolumn{2}{c}{Non-blocking} \\
                                     & Time & Energy                  & Time & Energy                    \\
        \hline
        29 & 0.51~s & 15.3~kJ & 0.53~s & 15.0~kJ \\
        30 & 0.59~s & 15.7~kJ & 0.74~s & 18.7~kJ \\
        31 & 0.80~s & 20.8~kJ & 0.97~s & 24.2~kJ \\
        32 & 9.63~s & 191~kJ  & 8.82~s & 179~kJ  \\
    \end{tabular}
    \caption{Time/energy per gate in the Hadamard benchmark on qubits~29--32 using blocking/non-blocking MPI.}
    \label{tab:hbench}
\end{table}

Fifty Hadamard gates were applied to a specific qubit, from 0~to 37, on 64~nodes and per-gate values calculated. \Cref{tab:hbench} presents the results on qubits~29--32. Up until qubit~29 the time per gate is roughly constant at 0.5~s, and the energy is approximately 15~kJ. At qubits 30~and 31~both values slightly increase, towards 1.0~s and 25~kJ due to data dependencies spreading to multiple NUMA regions. At qubit~32 and above, the runtime drastically rises to almost 10~s per gate for blocking communication, and 9~s for our non-blocking version, similarly, the energy shoots up to 190~kJ and 180~kJ respectively. The twenty-fold increase in runtime is caused by MPI, which now needs to communicate all the local elements of the statevector to another process. It is clear that non-blocking messages mitigate some of the communication costs.  
 
\begin{figure}[ht]
    \centering
    \includegraphics[width=.8\linewidth]{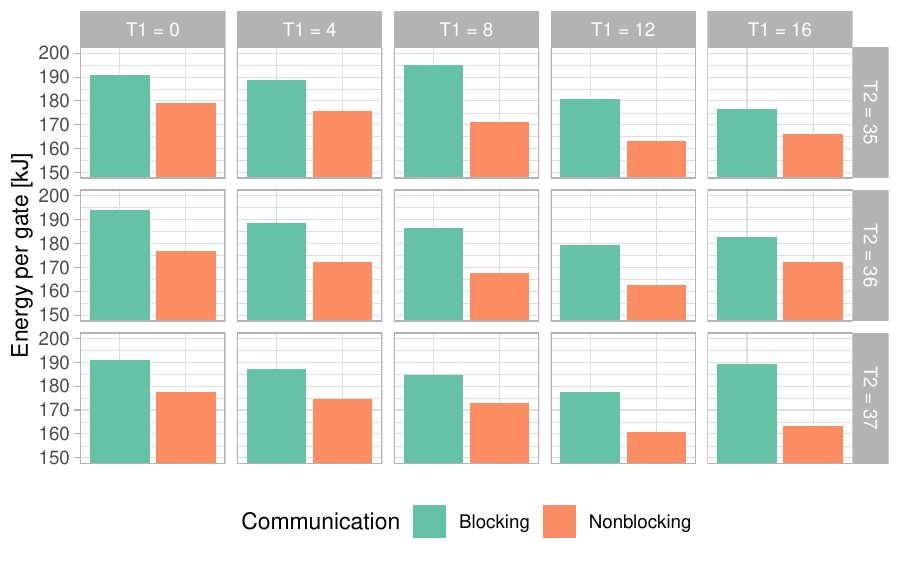}
    \caption{Energy consumption of the SWAP benchmark.}
    \label{fig:swapbench}
\end{figure}

Next, a similar benchmark with 50~SWAP gates was run. The SWAP gates require two targets, and as long as one is in the distributed regime, the operation entails communication. Since it is expensive to try all possible combinations of targets, we instead selected 5~local targets $[0, 4, 8, 12, 16]$, and 3 distributed targets $[35, 36, 36]$. The energy results are shown on \cref{fig:swapbench}. Again, the non-blocking communication visibly improved the performance -- with blocking communication the time per gate ranges from 9.0~s to 9.75~s and the energy consumption is from 180~kJ to 195~kJ; meanwhile, our non-blocking implementation requires 8.25~s to 9.0~s per gate and 160~kJ to 180~kJ.

\begin{figure}[ht]
    \centering
    \includegraphics[width=.8\linewidth]{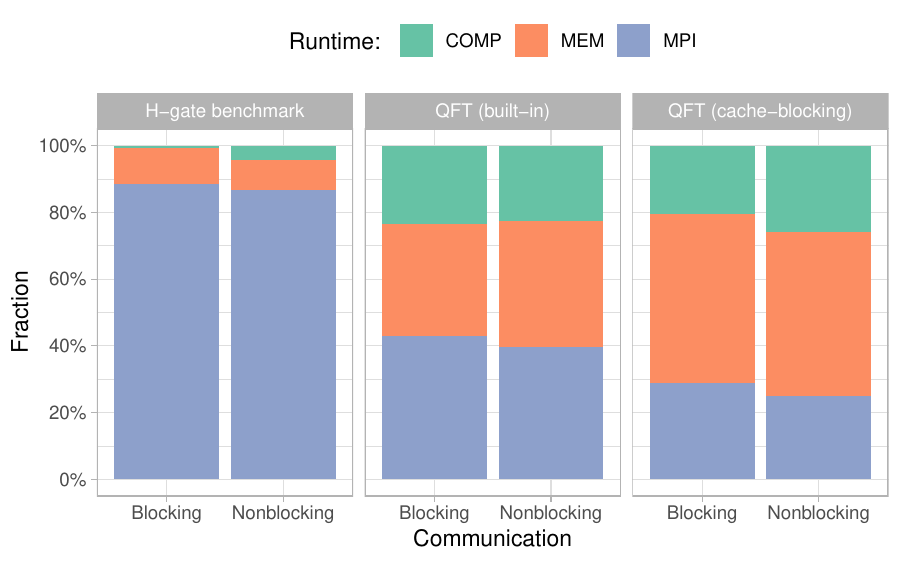}
    \caption{Profiles of the Hadamard and QFT benchmarks. }
    \label{fig:opt-profiles}
\end{figure}

Lastly, we profiled the different QFT implementations and compared them to the worst-case communication scenario (last qubit Hadamard benchmark). The built-in QFT in QuEST implements the circuit shown in \cref{fig:qft-standard}, but the controlled phase gates are applied more efficiently. The cache blocking circuit is as \cref{fig:qft-cache-blocking}, but the swaps are done after the 30th Hadamard gate to prevent any increase in gate execution time, and the controlled phase gates used the optimised QuEST implementation. 

The profiles of each run are presented on \cref{fig:opt-profiles}. In the Hadamard benchmark MPI completely dominates the runtime. The QFT gates are mostly local, so communication only takes up to 43\%~of runtime, and the rest is split roughly 2:1~between memory access and computation. By applying our optimisation, we managed to reduce communication to~25\%.

\subsection{Best QFT Performance}

Our final experiments were runs designed to use a significant share of ARCHER2 (which has 5,860 nodes), with and without optimisations. We tested the QFT on 43 and 44~qubits, using 2,048 and 4,096~nodes respectively. The `Fast' version uses cache-blocking to move all the Hadamard gates out of the distributed regime, and implements non-blocking communication for the SWAP gates. Results are shown in \cref{tab:qft-big-runs}.

\begin{table}[ht]
    \centering
    \begin{tabular}{l | l l l l}
        Experiment  & Built-in  & Fast    & Built-in  & Fast    \\
        \hline
        Qubits      & 43        & 43      & 44        & 44      \\
        Nodes       & 2048      & 2048    & 4096      & 4096    \\
        \hline
        Runtime     & 417~s     & 270~s   & 476~s     & 285~s   \\
        Energy      & 294~MJ    & 206~MJ  & 664~MJ    & 431~MJ  \\
    \end{tabular}
    \caption{Runtime and energy consumption of large QFT runs on ARCHER2. `Built-in' is using the standard QuEST QFT, `Fast' is our modified version. }
    \label{tab:qft-big-runs}
\end{table}

We achieved 35\% and 40\% improvements in runtime, along with 30\% and 35\% reductions in energy for the 2,048-node and 4,096-node runs respectively. The biggest energy improvement was 233~MJ, which is around 65~kWh saved in a little more than 3~minutes.

\section{Conclusions and Future Work}

We explored multiple ways to mitigate the energy consumption on ARCHER2 with QuEST framework. First, we highlighted the importance of using moderate CPU frequencies, which reduces the energy consumption with only a small runtime penalty. We note that the benefits end at 2.00~GHz -- any further decrease worsens the runtime while keeping the energy usage fixed. Second, we investigated whether high memory nodes are beneficial. Results were mixed, although high memory nodes are more efficient for the CU budget, the energy consumption is sometimes slightly higher and other times slightly lower than for the standard nodes. In particular, due to memory bandwidth being a limiting factor, we do not recommend specifying high-memory nodes over more balanced nodes if selecting hardware for this application. Last, we explored transpiling the simulated circuit to be \textit{cache-blocking}, which can reduce the number of distributed operations. The QFT is a perfect circuit for cache blocking, as SWAP gates are already a part of it. We also found that communication in QuEST could be improved on high-end networks using non-blocking communication. Applying all these optimisations, we managed to simulate 44~qubit QFT 40\%~faster, saving 35\%~energy. This work provides energy benchmarks both for both classical simulation codes and real quantum hardware to consider.

In future work we will further explore cache-blocking. If SWAP gates are the only distributed operations, communication could potentially be halved, as swapping only modifies half of the statevector. With this improvement, ARCHER2 could possibly simulate up to 45~qubits. It would also be useful to implement a cache-blocking transpiler, for example as an LLVM optimisation pass using the newly developed QIR~\cite{QIR}. We believe it would be beneficial to reimplement QuEST's core data-structures using a complex data type rather than separate real and imaginary arrays, in order to improve data locality. Finally, we will explore the impact on performance and energy usage of porting QuEST to multiple GPUs.

\begin{acks}

This work used the ARCHER2 UK National Supercomputing Service (\url{https://www.archer2.ac.uk}).

\end{acks}

\bibliographystyle{ACM-Reference-Format}  
\bibliography{refs}

\end{document}